\documentclass[conference]{IEEEtran}
\IEEEoverridecommandlockouts
\usepackage{cite}
\usepackage{amsmath,amssymb,amsfonts}
\usepackage{graphicx}
\usepackage{textcomp}
\usepackage{xcolor}
\def\BibTeX{{\rm B\kern-.05em{\sc i\kern-.025em b}\kern-.08em
    T\kern-.1667em\lower.7ex\hbox{E}\kern-.125emX}}

\usepackage[T1]{fontenc}
\usepackage{balance}
\usepackage{hyperref}

\usepackage[autostyle=false, style=english]{csquotes}
\MakeOuterQuote{"}

\usepackage{algorithm}
\usepackage{algorithmicx}
\usepackage{algpseudocode}

\usepackage{multirow}
\usepackage{xspace}

\ifCLASSOPTIONcompsoc
\usepackage[caption=false,font=normalsize,labelfont=sf,textfont=sf]{subfig}
\else
\usepackage[caption=false,font=footnotesize]{subfig}
\fi

\usepackage[export]{adjustbox} %

\usepackage{tikz}
\usepackage{textcomp}

\newcommand{\altparagraph}[1]{%
  \smallskip
  \textbf{#1}
  \noindent
}

\newcommand{\CommentLine}[1]{
    \State // \texttt{#1}
}

\def\name{\text{\textsc{Hiku}}\xspace}

\definecolor{darkblue}{HTML}{09357A}

\newcommand\copyrighttext{
  \footnotesize \textcopyright~2025 IEEE. Personal use of this material is permitted. Permission from IEEE must be obtained for all other uses, in any current or future media, including reprinting/republishing this material for advertising or promotional purposes, creating new collective works, for resale or redistribution to servers or lists, or reuse of any copyrighted component of this work in other works. This article is published in the 2025 IEEE 25th International Symposium on Cluster, Cloud and Internet Computing (CCGrid). DOI: \href{https://doi.org/10.1109/CCGRID64434.2025.00034}{\textcolor{darkblue}{10.1109/CCGRID64434.2025.00034}}.}
\newcommand\copyrightnotice{
\begin{tikzpicture}[remember picture,overlay]
\node[anchor=south,yshift=10pt] at (current page.south) {\fbox{\parbox{\dimexpr\textwidth-\fboxsep-\fboxrule\relax}{\copyrighttext}}};
\end{tikzpicture}
}

\begin{document}

\title{\name: Pull-Based Scheduling for Serverless Computing}

\author{
    \IEEEauthorblockN{Saman Akbari\IEEEauthorrefmark{1}, Manfred Hauswirth\IEEEauthorrefmark{1}\IEEEauthorrefmark{2}}
    \IEEEauthorblockA{\IEEEauthorrefmark{1}Technische Universität Berlin, Open Distributed Systems, Berlin, Germany}
    \IEEEauthorblockA{\IEEEauthorrefmark{2}Fraunhofer Institute for Open Communication Systems (FOKUS), Berlin, Germany \\
                      Email: \texttt{\{akbari, manfred.hauswirth\}@tu-berlin.de}}
}

\maketitle
\copyrightnotice

\begin{abstract}
Serverless computing promises convenient abstractions for developing and deploying functions that execute in response to events. In such Function-as-a-Service (FaaS) platforms, scheduling is an integral task, but current scheduling algorithms often struggle with maintaining balanced loads, minimizing cold starts, and adapting to commonly occurring bursty workloads.
In this work, we propose pull-based scheduling as a novel scheduling algorithm for serverless computing. Our key idea is to decouple worker selection from task assignment, with idle workers requesting new tasks proactively.
Experimental evaluation on an open-source FaaS platform shows that pull-based scheduling, compared to other existing scheduling algorithms, significantly improves the performance and load balancing of serverless workloads, especially under high concurrency. The proposed algorithm improves response latencies by 14.9\% compared to hash-based scheduling, reduces the frequency of cold starts from 43\% to 30\%, increases throughput by 8.3\%, and achieves a more even load distribution by 12.9\% measured by the requests assigned per worker.

\end{abstract}


\begin{figure*}[!th]
    \centering
    \includegraphics[width=\textwidth]{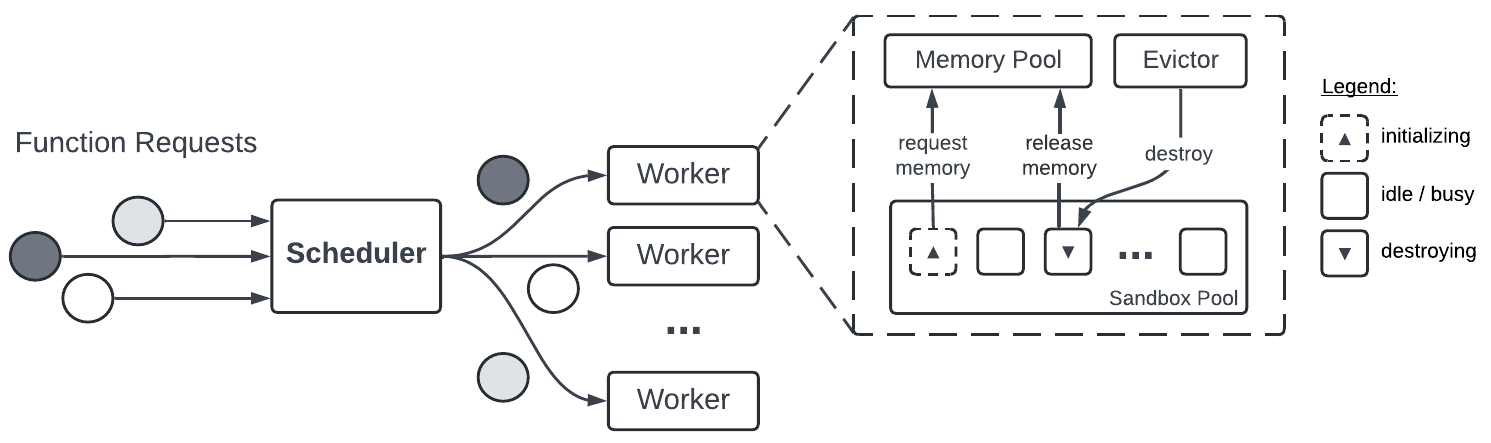}
    \caption{Architectural overview of function-as-a-service platforms. Workers manage a memory pool and execute functions in response to events in virtualized execution environments (sandboxes). The scheduler assigns function requests to available workers for execution.}
    \label{fig:faas_platform_architecture}
\end{figure*}

\section{Introduction}
\label{sec:introduction}

Function-as-a-Service (FaaS) is a cloud computing model that lets developers run functions without managing infrastructure tasks such as provisioning, auto-scaling, or scheduling. This provides a "serverless" experience that simplifies application development. All major cloud providers offer FaaS solutions, such as AWS Lambda~\cite{aws_lambda}, Google Cloud Functions~\cite{google_cloud_functions}, or Microsoft Azure Functions~\cite{microsoft_azure_functions}, along with several open-source FaaS platforms~\cite{hendrickson2016serverless, openwhisk, openfaas, knative}. The applications of FaaS are widespread and range from simple event-driven applications~\cite{ali2020batch} to complex workflows~\cite{fouladi2019laptop, li2022funcx, mahgoub2022wisefuse}.

Delivering FaaS functionalities presents significant operational challenges, because public clouds operate at a very large scale, e.g., processing nearly 1 billion function calls per day~\cite{zhangFasterCheaperServerless2021}, are multi-tenant, and serve functions with varying invocation patterns and different resource requirements.
A key element of any FaaS platform is the scheduler, which is responsible for assigning tasks to available workers to execute functions. Effective scheduling is critical to ensure consistent and high performance, particularly amidst fluctuating demands and heterogeneous workloads.

Serverless functions suffer from cold starts: Initializing a new instance of a function involves starting an execution environment, such as a virtual machine, fetching function code, starting a language runtime, and installing dependencies~\cite{silva2020prebaking, fuerst2022locality, shahrad2020serverless}. This results in significant delays when a function is executed without a "pre-warmed" instance. To mitigate the cold start problem, FaaS platforms use a "keep-alive" strategy, where function instances remain in an idle state for a period of time after execution to recycle these instances and avoid re-initialization when handling subsequent requests~\cite{wang2018peeking}.

Traditional scheduling algorithms, like least connections or randomized methods~\cite{hellemans2018power}, are inadequate for serverless environments as they do not address cold starts.
Instead, schedulers on FaaS platforms such as OpenLambda~\cite{hendrickson2016serverless}, Knative~\cite{knative} and OpenFaaS~\cite{openfaas} typically use hashing to increase function locality. In this approach, the function type of an incoming request is the input to generate a fixed-size output value that corresponds to a worker~\cite{fuerst2022locality, abdi2023palette, aumala2019beyond}. Function locality ensures that a specific function type is consistently assigned to the same worker, which reduces the number of cold starts and improves performance by reusing idle function instances. Although balancing loads across workers is a fundamental goal of scheduling, hash-based approaches can only do so under ideal conditions---specifically, when invocation patterns for all functions follow a uniform distribution. In practice, the situation is quite different: function invocations typically show a heavily skewed distribution, where a small subset of functions receives the majority of requests~\cite{zhangFasterCheaperServerless2021}, leading to load imbalances and slower executions.

In this paper, we propose a novel pull-based scheduling algorithm for serverless computing: \name (Japanese for "pull"), which achieves high function locality, balances worker loads, and scales efficiently as concurrency increases. Our key idea is to decouple worker selection from task assignment: After finishing the execution of a function, workers proactively request new tasks by enqueuing in idle queues. Our algorithm belongs to the class of Join-Idle-Queue algorithms~\cite{lu2011join}, which have proven effective in distributed environments~\cite{wang2018distributed}, but have not yet been applied to serverless computing.

Pull-based scheduling offers several advantages in serverless environments: First, workers dynamically request tasks based on their current capacity and naturally balance loads. Second, pull-based scheduling does not rely on a centralized scheduler. Third, in distributed systems with multiple schedulers, we reduce the need for synchronization, unlike traditional push-based scheduling algorithms that either try to maintain an often inaccurate and costly global view of worker states, or are oblivious to worker states. Finally, pull-based scheduling is quite simple to implement.

Our research contributes to research in serverless computing in several ways: We analyze real-world traces of a commercial FaaS platform~\cite{zhangFasterCheaperServerless2021} and show that commonly used scheduling techniques, such as hash-based algorithms and random scheduling, are suboptimal for serverless workloads. Based on these findings, the paper introduces pull-based scheduling, a decentralized, locality and load-aware scheduler where idle workers proactively request tasks. We then implement pull-based scheduling on OpenLambda~\cite{hendrickson2016serverless}, a popular open-source FaaS platform, and demonstrate performance and load balancing improvements over existing scheduling algorithms.
Our experimental results show that pull-based scheduling effectively balances loads, reduces cold starts, improves response latencies, and increases throughput especially under high concurrency.

The rest of this paper is structured as follows: We first provide the necessary background information on serverless computing and hashing (Section~\ref{sec:backgrund}). Next, we formalize scheduling in FaaS platforms and discuss its challenges (Section~\ref{sec:scheduling_in_faas_platforms}), which motivate the design and implementation of pull-based scheduling (Section~\ref{sec:pull_based_scheduling}). We then evaluate the performance of pull-based scheduling on OpenLambda (Section~\ref{sec:evaluation}), discuss related work in Section~\ref{sec:related_work}, and finally conclude the paper (Section~\ref{sec:conclusion}). The implementation of pull-based scheduling and all artifacts produced are openly accessible to foster replicability (Section~\ref{sec:open_data}).

\section{Background}
\label{sec:backgrund}
We start by outlining the architecture of FaaS platforms and the role of scheduling. Then, we elaborate how the lifecycle of functions significantly affects scheduling algorithms in serverless computing. Finally, we describe consistent hashing, which is a common scheduling algorithm in FaaS platforms, and explain its limitations that motivate our research.

\subsection{Architecture of Function-as-a-Service Platforms}
Figure~\ref{fig:faas_platform_architecture} shows the main components of a typical FaaS platform. At their core, FaaS platforms consist of a set of \textit{workers}, a set of \textit{functions}, and a \textit{scheduler}.
Requests follow an event-driven programming model in which workers execute stateless functions in response to specific triggers, such as HTTP requests, scheduled events, or messages from other services. On invocation, workers execute functions in a virtualized environment. They manage a memory pool to allocate resources to these instances. To optimize resource usage and free resources for new invocations, workers have an evictor component that removes idle function instances.
FaaS schedulers assign incoming requests to available workers, which process these requests and finally return results.

\subsection{Function Lifecycle}
\label{sec:function_lifecycle}
\begin{figure}[!th]
    \centering
    \includegraphics[width=\columnwidth]{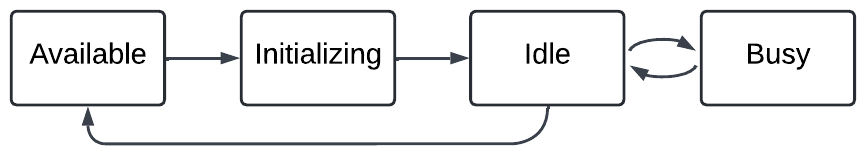}
    \caption{Lifecycle of function instances. Cold starts occur when there is no function instance for the requested function type in an idle state on a worker.}
    \label{fig:function_lifecycle}
\end{figure}

FaaS platforms keep function instances in an idle state for a period of time after execution to recycle these instances and avoid re-initialization when handling subsequent requests~\cite{wang2018peeking}. When a new request comes in and there is no idle instance for the requested function type, workers trigger a cold start to create a new instance of that function. Figure~\ref{fig:function_lifecycle} shows the function lifecycle: From the initial \textit{available} state, functions transition through \textit{initializing}, \textit{idle}, and \textit{busy} states upon invocation, then again enter the \textit{idle} state after execution. Once an instance is initialized, it can only execute requests of the same function type. After a period of inactivity, instances time out and return to the \textit{available} state.

Cold starts are a big challenge in FaaS platforms, because they slow down requests significantly. Table~\ref{tab:response_latencies} shows average cold start and warm start times over 20 runs for functions in the FunctionBench suite~\cite{kimFunctionbenchSuiteWorkloads2019} on an OpenLambda~\cite{hendrickson2016serverless} worker running on a \texttt{m5.xlarge} AWS EC2 virtual machine. On average, cold start executions are 1.79$\times$ slower than warm start executions.

\begin{table}[!th]
\centering
\caption{Average response latencies of applications in FunctionBench~\cite{kimFunctionbenchSuiteWorkloads2019}. Cold starts significantly slow down executions.}
\resizebox{0.9\columnwidth}{!}{ %
\begin{tabular}{lll}
\hline
Application & Cold Start (ms) & Warm Start (ms) \\ \hline
chameleon            & 536                   & 392                  \\
dd                   & 706                   & 549                  \\
float\_operation     & 263                   & 94                   \\
gzip\_compression    & 510                   & 303                  \\
json\_dumps\_loads   & 269                   & 105                  \\
linpack              & 282                   & 58                   \\
matmul               & 284                   & 125                  \\
pyaes                & 329                   & 149                  \\ \hline
\end{tabular}
}
\label{tab:response_latencies}
\end{table}

The impact of cold starts on performance has led to various optimization strategies in FaaS platforms, either by reducing the \textit{frequency} of cold starts~\cite{agarwal2021reinforcement, fuerst2022locality, kim2021scheduling} or by reducing the \textit{impact} of cold starts~\cite{roy2022icebreaker, aumala2019beyond}. FaaS schedulers typically aim to reduce the frequency of cold starts by maximizing function locality, i.e., running repeated invocations of a function on the same worker.

\subsection{Consistent Hashing}
Schedulers on FaaS platforms such as OpenLambda~\cite{hendrickson2016serverless}, Knative~\cite{knative} and OpenFaaS~\cite{openfaas} typically use hashing to increase function locality, where the function type of an incoming request is the input to generate a fixed-size output value that corresponds to a worker~\cite{fuerst2022locality, abdi2023palette, aumala2019beyond}. A na\"ive scheduler would partition the hash to a worker by taking the modulo with the total number of workers. However, adding or removing workers during auto-scaling changes the modulus. As a result, auto-scaling would cause significant performance degradation, because potentially many keys may need to be redistributed.

\begin{figure}[!th]
    \centering
    \includegraphics[width=0.7\columnwidth]{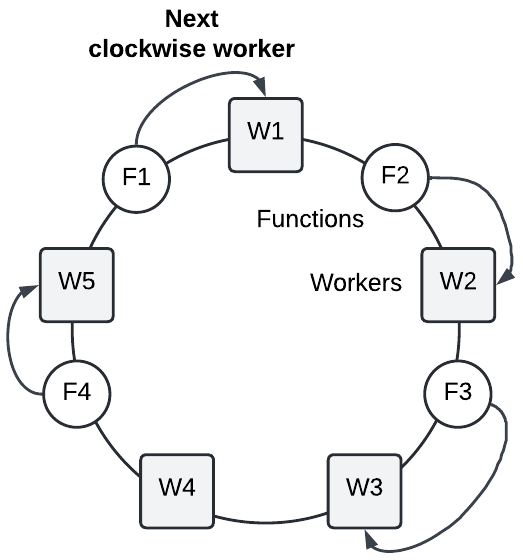}
    \caption{Consistent hashing assigns function types (keys) to the next clockwise worker (value) on the hash ring. Adding or removing workers requires only minimal redistribution of keys.}
    \label{fig:consistent_hashing}
\end{figure}

A common partitioning algorithm in FaaS platforms to minimize key redistribution is consistent hashing, which conceptually places function types (keys) and workers (values) on a "hash ring" (see Figure~\ref{fig:consistent_hashing}). Consistent hashing assigns function requests to the next worker clockwise on the hash ring. Adding or removing a worker affects \textit{only} the function types assigned to that specific worker on the hash ring, as they are reassigned to the next (different) worker in the clockwise direction. This reduces the performance impact on FaaS platforms by minimizing the number of function types that need to be redistributed during auto-scaling of workers.

Consistent hashing has limitations in real-world scenarios. An extension, consistent hashing with bounded loads (CH-BL)~\cite{mirrokni2018consistent}, addresses the issue of finite worker capacity by introducing a load threshold parameter. When the load of a worker exceeds this threshold, it assigns new requests to the next non-overloaded worker clockwise on the hash ring. However, CH-BL can lead to cascaded overflows: In high-load scenarios, the next clockwise workers are likely to become sequentially overloaded. Another extension, random jumps for consistent hashing (RJ-CH)~\cite{chen2021revisiting}, addresses cascaded overflows by randomly selecting a non-overloaded worker when the next clockwise worker is at capacity, but at the expense of function locality.

\section{Scheduling in Function-as-a-Service Platforms}
\label{sec:scheduling_in_faas_platforms}
In the following, we formalize scheduling in FaaS platforms and elaborate on the associated challenges, which we refer to as the "serverless scheduling trilemma."

\subsection{Problem Definition}

\altparagraph{System:}
We can formally define scheduling in FaaS platforms as follows:

\begin{itemize}
    \item $F = \{f_1, f_2, \dots, f_k\}$ is the set of available functions.
    \item $W = \{w_1, w_2, \dots, w_m\}$ is the set of available workers.
    \item $R = (\{r_1, r_2, \dots, r_n\}, \leq)$ is the totally ordered sequence of function requests, ordered by the arrival time of the requests $t_{arrival}(r_1) \leq t_{arrival}(r_2) \leq \dots \leq t_{arrival}(r_n)$.
\end{itemize}

\altparagraph{Requests:}
Each request $r_i \in R$ is characterized by:

\begin{itemize}
    \item $f(r_i) \in F$ is the type of function requested.
    \item $mem(r_i)$ is the memory allocated for the request.
    \item $t_{arrival}(r_i)$ is the arrival time of the request.
\end{itemize}

\altparagraph{Workers:}
Each worker $w_j \in W$ is characterized by:

\begin{itemize}
    \item $cap(w_j)$ is the memory capacity of $w_j$.
    \item $usage(w_j, t)$ is the memory usage of $w_j$ at time $t$.
    \item $I_{w_j, t} \subseteq F$ is the set of idle function instances on $w_j$ at time $t$.
\end{itemize}

\altparagraph{Function Execution:}
The lifecycle of a function execution on a worker is characterized by:

\begin{itemize}
    \item If $f(r_i) \notin I_{w_j, t_{arrival}(r_i)}$, the worker initializes a new function instance (cold start).
    \item The function is then executed.
    \item After execution, the function instance remains idle for a period of time $t_{idle}$.
    \item An initialized function instance can only execute requests of the same type, i.e., $f(r_i) = f(r_j)$ for $r_i, r_j \in R, i \neq j$.
    \item If the instance remains idle for longer than $t_{idle}$, the worker evicts it. Idle instances are force-evicted if $usage(w_j, t)$ exceeds $cap(w_j)$.
\end{itemize}

\altparagraph{Scheduling:}
Scheduling is an online algorithm that finds a mapping $S : R \rightarrow W \times \mathbb{R}^+$ for each request $r_i$:

\begin{equation}
S(r_i) = (w_j, t_{exec})
\end{equation}

Where:

\begin{itemize}
    \item $w_j \in W$ is the worker assigned to execute the request.
    \item $t_{exec} \geq t_{arrival}(r_i)$ is the time at which the execution begins.
\end{itemize}

\subsection{Challenges}
\label{sec:faas_scheduling_challenges}
This section discusses four key challenges that schedulers face on FaaS platforms: multi-tenancy, skewed function popularity, heterogeneous function performance, and bursty invocations. We give empirical examples of these challenges using real-world traces from a commercial FaaS platform with the Azure Functions dataset~\cite{zhangFasterCheaperServerless2021}.

\altparagraph{Multi-tenancy:}
On FaaS platforms, multiple tenants, i.e., users, share infrastructure resources,
which can lead to resource contention and introduces complexities in resource allocation and performance isolation.
The challenge of multi-tenancy extends beyond just inter-tenant concerns: Even within the environment of a single tenant, different function types compete for resources and potentially impact each other's performance.

\begin{figure}[!th]
    \centering
    \includegraphics[width=0.8\columnwidth]{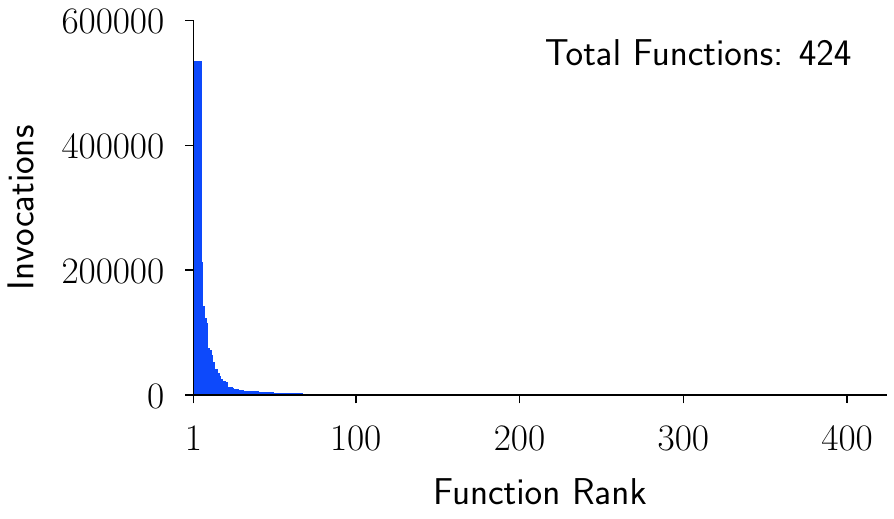}
    \caption{Skewed function popularity. Function invocations show a heavily skewed distribution, where a minority of functions receive the majority of requests.}
    \label{fig:function_popularity}
\end{figure}

\altparagraph{Skewed function popularity:}
Function invocations in FaaS platforms often show a heavily skewed distribution, where only a few functions receive the majority of requests (Figure~\ref{fig:function_popularity}). The top 10\% of popular functions in the Azure Functions dataset account for 92.3\% of all invocations, whereas the top 1\% alone account for 51.3\%.
This skewness poses a significant challenge for schedulers, where highly popular functions can cause resource starvation for less popular functions.

\begin{figure}[!th]
    \centering
    \includegraphics[width=0.8\columnwidth]{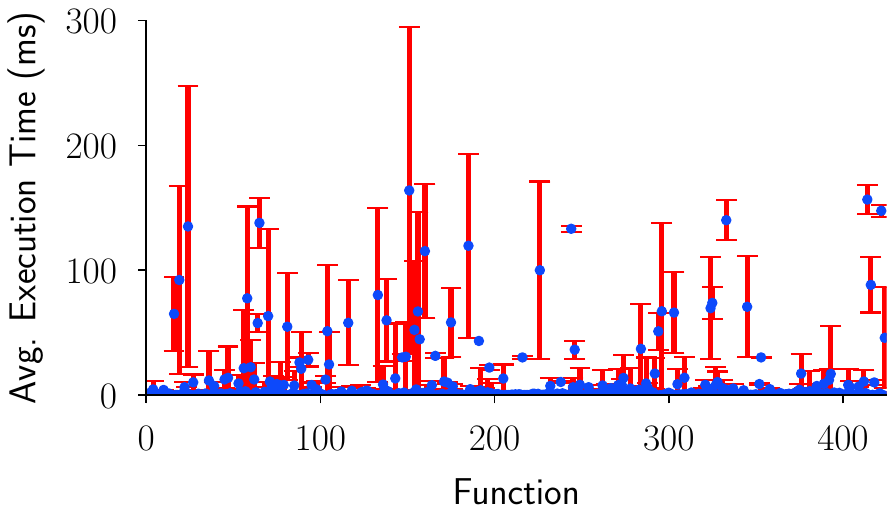}
    \caption{Heterogeneous function performance. The execution time of functions varies significantly both between different functions and within the same function. Error bars represent the standard deviation of execution times, and functions are ordered by their first appearance in the dataset.}
    \label{fig:heterogeneous_function_performance}
\end{figure}

\altparagraph{Heterogeneous function performance:}
Repeated executions of the same function show significant variations in execution time, e.g., due to performance fluctuations in the cloud~\cite{schirmer2023night} or input-dependent control flow paths (Figure~\ref{fig:heterogeneous_function_performance}).
This makes it difficult to maintain stable performance and achieve optimal resource utilization.

\begin{figure}[!th]
    \centering
    \includegraphics[width=0.8\columnwidth]{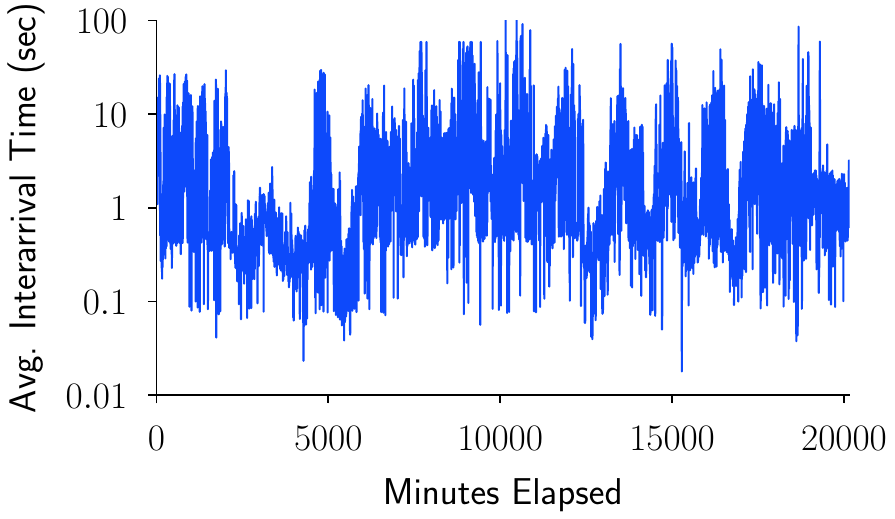}
    \caption{Bursty invocations. Invocations on FaaS platforms often occur in sudden bursts, as the average interarrival time per minute changes rapidly.}
    \label{fig:bursty_invocations}
\end{figure}

\altparagraph{Bursty invocations:}
Workloads on FaaS platforms are often characterized by sudden bursts of incoming requests, with interarrival times increasing or decreasing by up to 13.5$\times$ within a minute in the Azure Functions dataset (Figure~\ref{fig:bursty_invocations}).
Schedulers should strive to adapt to these sudden changes in workload intensity.

\subsection{The Serverless Scheduling Trilemma}
Schedulers in serverless computing face a complex set of goals, three of which stand out: load balancing, latency minimization, and performance stability.

\textit{Load balancing} is critical to efficient resource utilization. By evenly distributing workloads across available workers, schedulers prevent individual workers from becoming overloaded, leading to better overall resource efficiency. \textit{Minimizing latency} is equally important, as it directly impacts the responsiveness of serverless functions. A common strategy for reducing latency is maximizing function locality, i.e., trying to ensure that function requests are scheduled on workers where the function is already warm and initialized to minimize the number of cold starts. \textit{Stable performance} is the third goal, particularly in multi-tenant environments and under bursty workloads. Achieving stability involves maintaining predictable and reliable performance under various conditions, thus upholding service level agreements.

However, we face a trilemma here, as optimizing for any goal tends to compromise the other two: Evenly distributing loads among workers can break function locality and therefore cause performance instability. Prioritizing function locality to minimize latency can lead to uneven load distribution and performance fluctuations. Ensuring performance stability might require resource allocation that hinders both optimal load balancing and function locality. Reaching a balance among these goals is a complex challenge for schedulers in serverless computing.

\begin{figure}[!t]
    \centering
    \includegraphics[width=0.9\columnwidth,left]{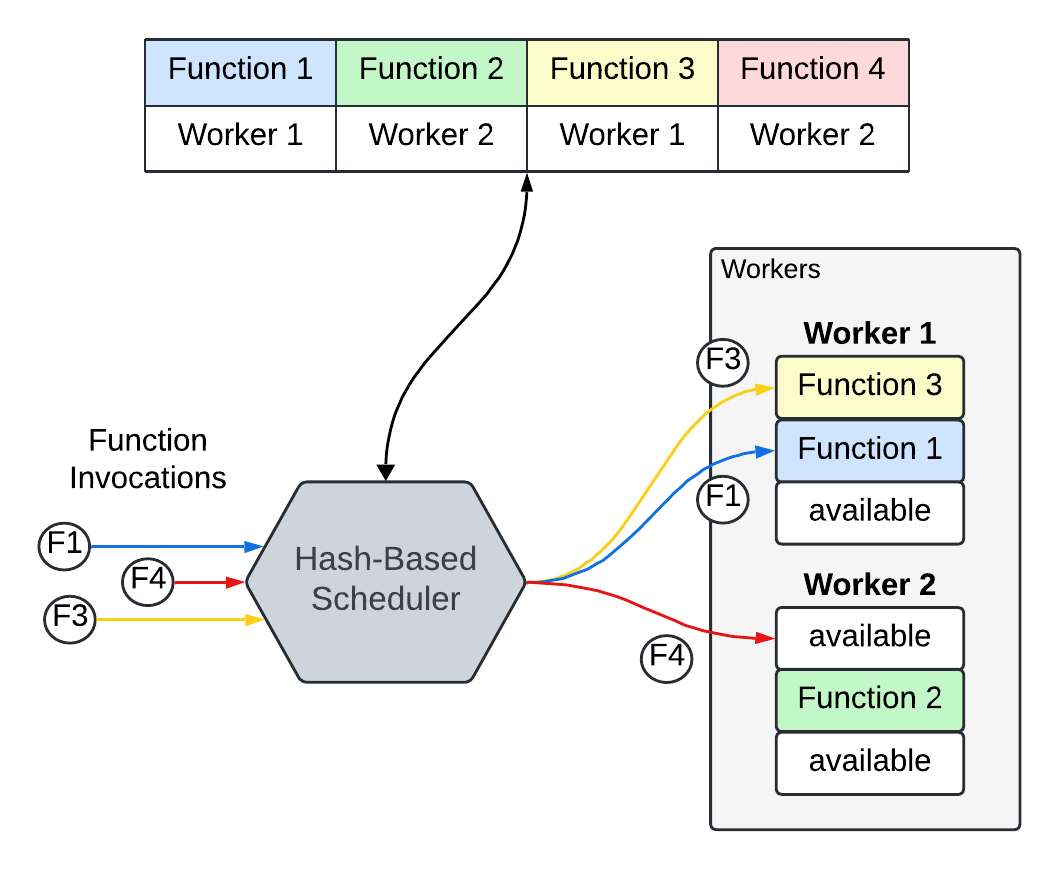}
    \caption{Hash-based scheduling. The scheduler assigns incoming tasks to workers based on a hash of the function type of the request (key) which maps to the corresponding worker (value).}
    \label{fig:consistent_hashing_scheduling}
\end{figure}

Existing scheduling algorithms for serverless computing often use \textit{consistent hashing} (CH) to improve function locality~\cite{fuerst2022locality, abdi2023palette, aumala2019beyond}. We show an overview of a hash-based scheduler in Figure~\ref{fig:consistent_hashing_scheduling} using an example. Although CH increases function locality and reduces latency, it can also result in load imbalances because of skewed function popularity and diverse performance characteristics. Due to these load imbalances, CH struggles with resource contention, which leads to increased latency as resources compete on frequently used workers, particularly in the multi-tenant settings of FaaS platforms or during peak invocation periods. Additionally, CH treats all functions uniformly and does not consider varying resource demands, which potentially creates bottlenecks, especially pronounced in heterogeneous workloads. These limitations highlight the need for more advanced scheduling approaches for serverless computing.

\section{Pull-Based Scheduling}
\label{sec:pull_based_scheduling}
To alleviate the above trilemma, we propose \textit{pull-based scheduling}, a novel scheduling algorithm for serverless computing. The core idea behind pull-based scheduling is to decouple worker selection from task assignment. We argue that by allowing idle workers to actively request new tasks, FaaS platforms can achieve high function locality, balanced loads, \textit{and} stable performance. This is a fundamental shift in authority compared to previous work on FaaS scheduling, which follows a \textit{push-based} model where the scheduler tries to select appropriate workers for incoming function requests.

\begin{figure}[!t]
    \centering
    \includegraphics[width=\columnwidth,left]{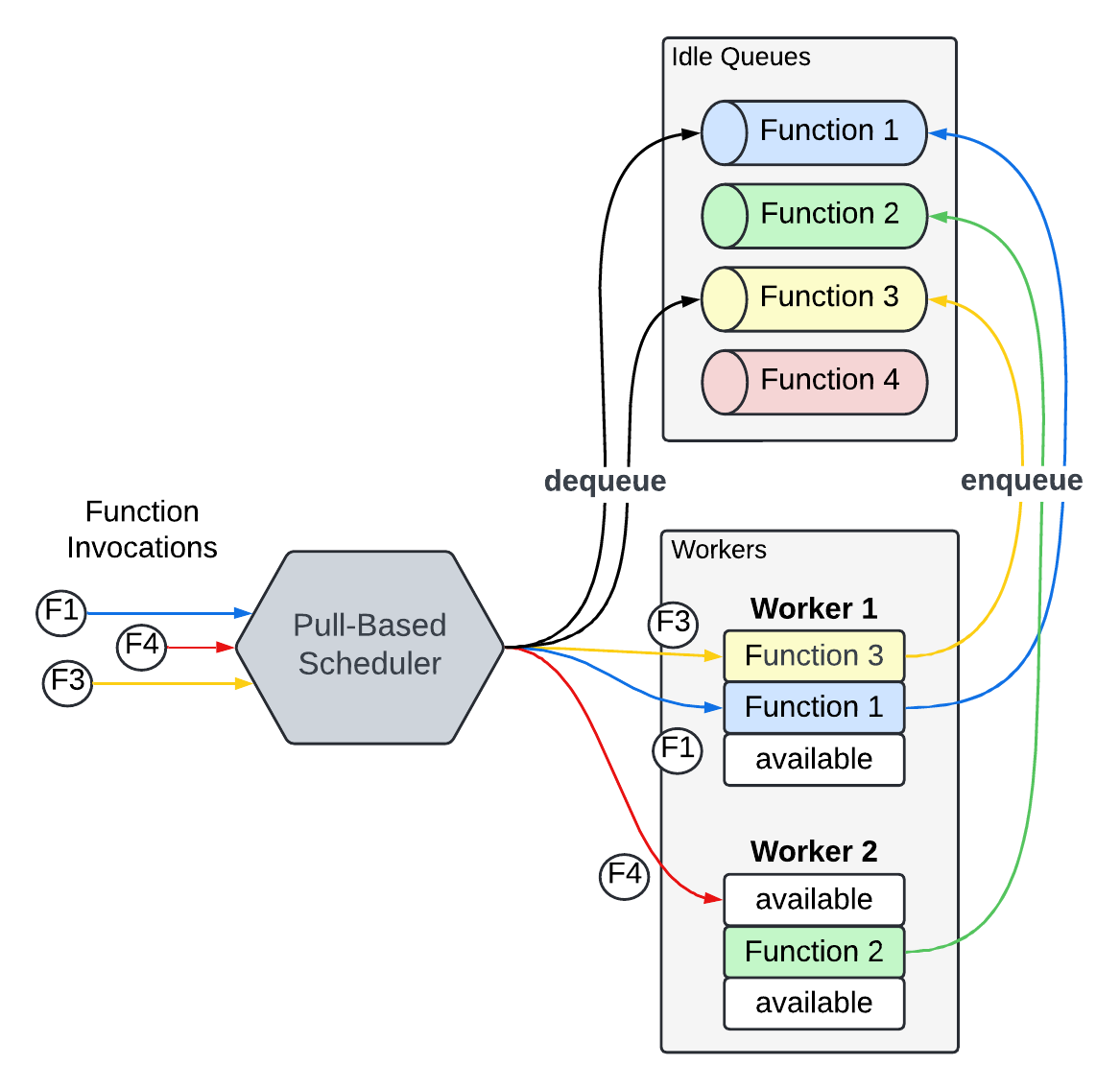}
    \caption{Pull-based scheduling. Workers with idle function instances indicate their readiness for new tasks by enqueuing in the idle queue of the previously executed function, which is sorted by the number of active connections. Pull-based scheduling dequeues from idle queues on invocation, if not empty; otherwise, it assigns the task to the least loaded worker.}
    \label{fig:pull_based_scheduling}
\end{figure}

Figure~\ref{fig:pull_based_scheduling} shows the design of our proposed algorithm. After executing a function, workers enqueue in the corresponding idle queue of the previous task to proactively signal their availability for new tasks, instead of passively waiting. When users invoke a function, pull-based scheduling first checks the idle queue of that function to assign the task to a worker with an idle (warm) function instance. If the idle queue is empty, the scheduler assigns the task to the least loaded worker.

We describe the details of pull-based scheduling in the rest of this section.

\begin{figure*}[!t]
    \centering
    \includegraphics[width=0.99\textwidth]{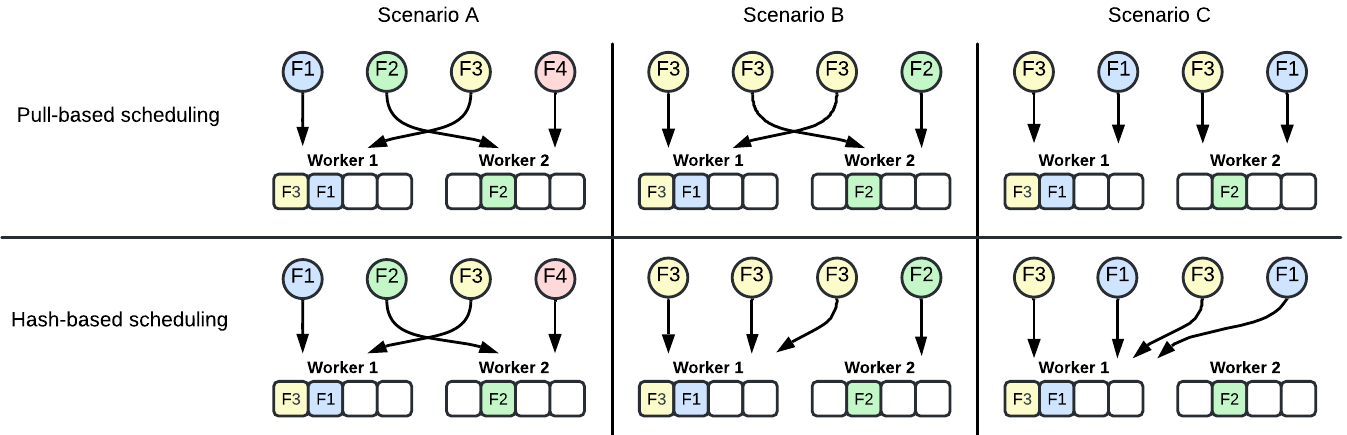}
    \caption{Three scheduling scenarios. Comparison of pull-based and hash-based scheduling with four function types (F1, F2, F3, F4) and two workers (W1, W2). W1 has idle instances of F1 and F3, whereas W2 has an idle instance of F2. In scenario A, uniform requests result in identical performance. In scenario B, skewed requests favoring F3 lead to better load balancing with pull-based scheduling. In scenario C with hash-based scheduling, W1 reaches its capacity, whereas pull-based keeps the load balanced. Overall, pull-based scheduling improves load balancing in these three scenarios while maintaining high function locality.}
    \label{fig:scheduling_scenarios}
\end{figure*}

\subsection{Pull Mechanism}
\label{sec:pull_mechanism}
The key principle of pull-based scheduling is the \textit{pull mechanism}, where workers with idle function instances pull requests for execution. Each function type $f$ has an idle (priority) queue $PQ_f$ that
sorts workers in ascending order by the number of active connections to prioritize the least loaded worker for a new task.
When a worker finishes executing a function, it does not passively wait for a new assignment, but \textit{proactively enqueues} in $PQ_f$ to signal readiness for new tasks. This mechanism inherently promotes function locality: Since workers enqueue in the idle queue corresponding to their last executed function type, subsequent requests for the same function are naturally directed to workers with warm instances, avoiding unnecessary cold starts.

When the scheduler selects a worker to execute a function of type $f$, we dequeue from the priority queue $PQ_f$.
We also incorporate a \textit{notification mechanism} for sandbox destruction. When a worker evicts a function instance, typically due to inactivity or resource reclamation, it notifies the scheduler. This notification allows the scheduler to update its state and remove the worker from the corresponding idle queue.

One might consider a simple alternative to our pull mechanism: a scheduler that continuously monitors worker states and assigns tasks to idle workers. However, the scheduler's view of these states can quickly become outdated, given that many serverless functions last only a few milliseconds (see Section~\ref{sec:scheduling_in_faas_platforms}). This issue is exacerbated in distributed systems with multiple schedulers that need to synchronize.
Our pull mechanism with proactive worker behavior enables a self-balancing system that handles serverless workloads efficiently, while minimizing the need for synchronization.

\subsection{Fallback Mechanism}
\label{sec:fallback_mechanism}
Pull-based scheduling incorporates a \textit{fallback mechanism} to ensure task assignment without delay when no worker has a warm function instance immediately available. In this scenario, we use least-connection scheduling: The scheduler counts the number of active connections for each worker and selects the worker with the fewest connections. If multiple workers have the same number of active connections, the scheduler randomly selects one of the tied workers. The fallback mechanism can be changed to other scheduling algorithms.

\begin{algorithm}[!th]
\caption{Pull-Based Scheduling}
\label{alg:pull_based_scheduling}
\begin{algorithmic}[1]
\Require Request $r$, Set of workers $W$, Priority queue of idle workers $PQ_f$ for function type $f$ of request $r$

\vspace{\baselineskip}
\State $f \gets$ function type of $r$ to execute
\If{$PQ_f$ is not empty}
    \CommentLine{Pull mechanism}
    \State $w \gets$ Dequeue worker $w$ from $PQ_f$
    \State Assign $r$ to $w$
\Else
    \CommentLine{Fallback mechanism}
    \State $L_{min} \gets \min_{w \in W} \text{Load}(w)$ \Comment{load expressed as, e.g., the number of active connections}
    \State $W_{min} \gets \{w \in W : \text{Load}(w) = L_{min}\}$
    \State $w \gets$ Random selection from $W_{min}$
    \State Assign $r$ to $w$
\EndIf

\vspace{\baselineskip}
\State \textbf{After execution:}
\If{$w$ finishes executing $r$}
    \State Enqueue $w$ in $PQ_f$
\EndIf

\vspace{\baselineskip}
\State \textbf{On eviction:}
\If{$w$ evicts function instance of type $f$}
    \State Remove first occurrence of $w$ from $PQ_f$
\EndIf

\vspace{\baselineskip}
\State \textbf{Note:} Priority queues $PQ_f$ are maintained sorted by $\text{Load}(w)$, e.g., the number of active connections.
\end{algorithmic}
\end{algorithm}

\subsection{Scheduling Scenarios}
\label{sec:scheduling_scenarios}
We compare how pull-based scheduling and hash-based scheduling perform in three different scheduling scenarios in Figure~\ref{fig:scheduling_scenarios}. Suppose that we have four function types (F1--F4) and two workers (W1, W2), each capable of executing up to four functions simultaneously. Worker 1 (W1) has one idle instance of F3 and F1 respectively from previous executions, whereas worker 2 (W2) has one idle instance of F2. The hash table of the hash-based scheduler assigns F1 and F3 to W1, whereas it assigns F2 and F4 to W2.

In the first scenario, requests for F1, F2, F3, and F4 arrive in sequence. Since the distribution of function types is uniform, both pull-based and hash-based scheduling perform identically. F1, F2, and F3 all result in a warm start due to the availability of idle instances, whereas F4 results in a cold start. The load is evenly balanced among the workers, with two functions running on W1 and two on W2.

In the second scenario, the function popularity is skewed, which is common for serverless workloads (see Section~\ref{sec:faas_scheduling_challenges}), with three requests for F3 and one for F2. Both pull-based scheduling and hash-based scheduling result in the same number of cold starts, because only one idle instance of F3 is available, but pull-based scheduling balances loads more evenly.

In the third scenario, requests for F3, F1, F3, and F1 arrive. With hash-based scheduling, the first requests for F3 and F1 result in warm starts, whereas the next two requests result in cold starts. Furthermore, W1 reaches its capacity. Pull-based scheduling also results in two warm starts and two cold starts, but the load is balanced between the two workers.

Both pull-based and hash-based scheduling achieve \textit{high function locality}, but hash-based scheduling tends to cause \textit{load imbalances} by consistently assigning requests for the same function to a single worker, even when that worker has no instances of idle functions.

\subsection{Implementation}
\label{sec:implementation}
We implemented pull-based scheduling on OpenLambda~\cite{hendrickson2016serverless} (commit \texttt{0a834ce}), an open-source FaaS platform. Our implementation extends an existing open-source scheduler for OpenLambda~\cite{aumala2019beyond} (commit \texttt{08303a1}) and is written in Go to maintain consistency with the existing codebase. We have open-sourced our implementation of pull-based scheduling to facilitate replicability of our experiments and further research (see Section~\ref{sec:open_data}).

\section{Evaluation}
\label{sec:evaluation}
In this section, we evaluate the performance of pull-based scheduling against three scheduling algorithms from the OpenLambda scheduler~\cite{aumala2019beyond}: least connections, random, and consistent hashing with bounded loads (CH-BL). For CH-BL, we set the load threshold parameter to the recommended value of 1.25~\cite{mirrokni2018consistent}.

\subsection{Experimental Setup}

\altparagraph{Environment:}
We deployed a cluster of 6 AWS EC2 \texttt{m5.xlarge} virtual machines with 16 GB of RAM, 4 vCPUs, and 150 GB EBS storage each. Five VMs host an OpenLambda worker and one VM hosts the scheduler.

\altparagraph{Workload:}
We selected 8 functions from FunctionBench~\cite{kimFunctionbenchSuiteWorkloads2019}, a popular benchmark suite for serverless computing (see Table~\ref{tab:benchmarks}). These benchmarks represent a wide range of serverless functions with CPU/memory-intensive, disk I/O-heavy, and network-bound operations.
To create a more realistic set of functions, we created 5 identical copies with unique names for each of the 8 functions for a total of 40 unique functions.

\begin{table}[!t]
\centering
\caption{Functions for evaluation from FunctionBench~\cite{kimFunctionbenchSuiteWorkloads2019}.}
\label{tab:benchmarks}
\resizebox{\columnwidth}{!}{
\begin{tabular}{lll}
\textbf{Type}                          & \textbf{Name}        & \textbf{Description}               \\ \hline
\multirow{5}{*}{\textit{CPU / Memory}} & chameleon            & String and text processing \\
                                       & float\_operation     & Floating-point arithmetic operations \\
                                       & linpack              & Dense linear equations \\
                                       & matmul               & Matrix multiplication \\
                                       & pyaes                & AES encryption algorithm \\ \hline
\multirow{2}{*}{\textit{Disk}}         & dd                   & File read and write \\
                                       & gzip\_compression    & File compression and decompression \\ \hline
\textit{Network}                       & json\_dumps\_loads   & JSON serialization and deserialization
\end{tabular}
}
\end{table}

\altparagraph{Execution:}
To generate requests, we used the k6 load testing tool~\cite{k6}.
We mimicked invocation patterns in the real-world invocation dataset of the commercial FaaS platform Azure Functions~\cite{zhangFasterCheaperServerless2021} using weighted random selection. Specifically, in each run, we randomly selected 40 functions from this dataset, calculated and normalized invocation probabilities, and then mapped these invocation probabilities to our functions.
Each invocation was followed by a sleep period of 0.1 to 1 second. We tested three different virtual user (VU) settings with 20, 50, and 100 VUs and ran each experiment for 5 minutes, evenly distributed across the three VU settings, and performed 20 runs per scheduling algorithm to account for performance fluctuations on cloud platforms.
For fairness, we seeded the random number generator in each run with the start date of the experiment so that the order of function invocations as well as sleep durations between invocations were identical for each scheduling algorithm.

\altparagraph{Metrics:}
We measured four metrics: response latency, throughput, cold start rate, and load imbalance.
Response latency and throughput are common metrics to evaluate the performance of a system,
whereas the cold start rate is specific to FaaS.
We define load imbalance as the coefficient of variation of the number of requests assigned per worker per second.
We also tested how well the scheduling algorithms handle different levels of concurrency with 20, 50, and 100 virtual users, measured as the processed requests per second.

\begin{figure*}[!th]
    \centering
    \begin{minipage}{0.49\textwidth}
        \centering
        \includegraphics[width=\textwidth]{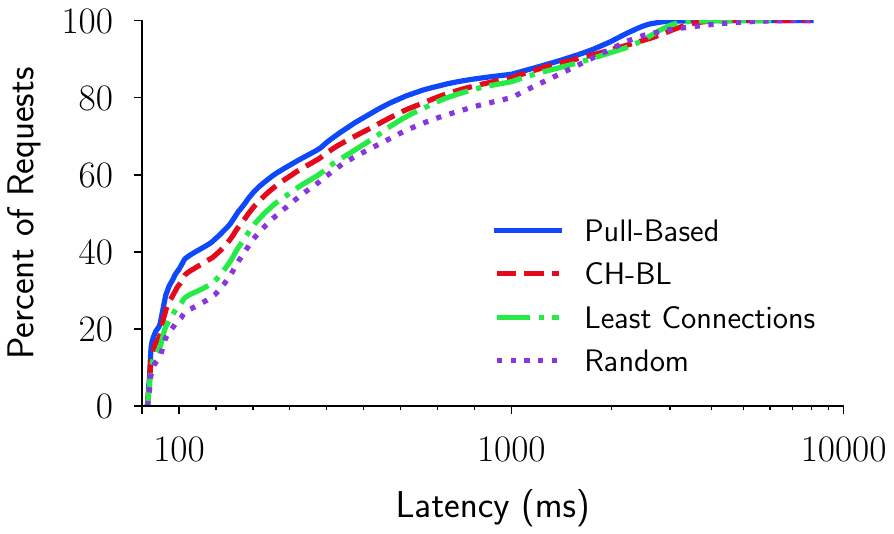}
        \caption{Response latencies. The cumulative distribution function for pull-based scheduling shows a noticeable shift to the left toward low latencies.}
        \label{fig:results_latency}
    \end{minipage}
    \hfill
    \begin{minipage}{0.49\textwidth}
        \centering
        \includegraphics[width=\textwidth]{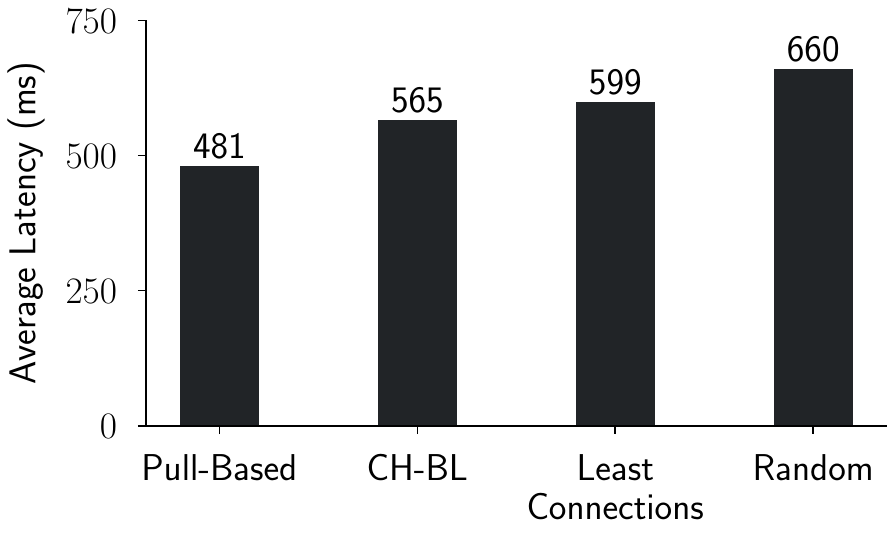}
        \caption{Average response latencies. Pull-based scheduling reduces response latencies by an average of 14.9\% to 27.1\%.}
        \label{fig:results_average_latency}
    \end{minipage}
\end{figure*}

\begin{figure*}[!th]
    \centering
    \begin{minipage}{0.49\textwidth}
        \centering
        \includegraphics[width=\columnwidth]{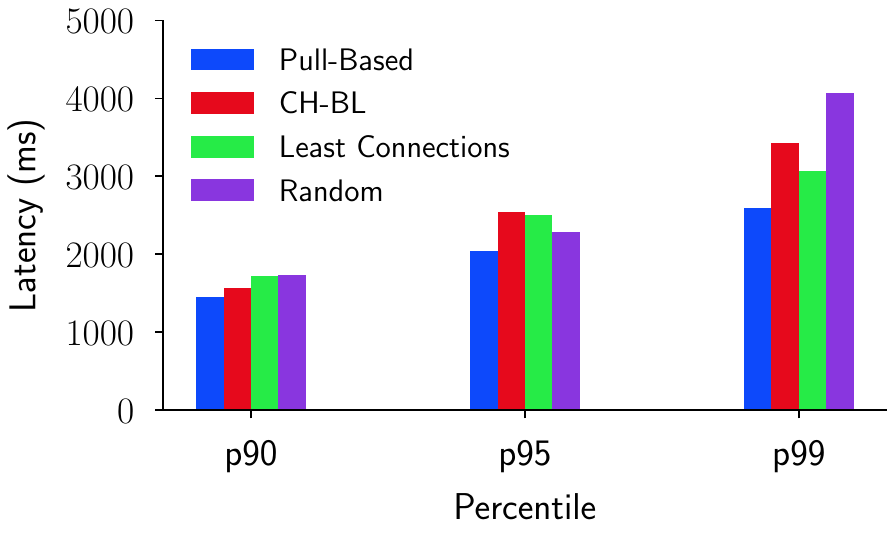}
        \caption{Tail latencies. Pull-based scheduling reduces tail latencies (90th, 95th, and 99th percentiles), particularly at the 99th percentile.}
        \label{fig:results_tail_latency}
    \end{minipage}
    \hfill
    \begin{minipage}{0.49\textwidth}
        \centering
        \includegraphics[width=\columnwidth]{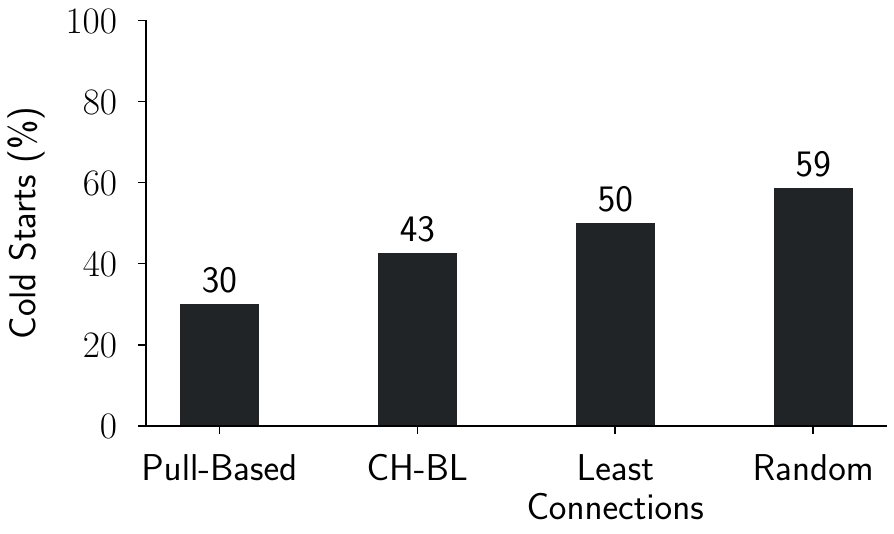}
        \caption{Cold starts. Pull-based scheduling significantly reduces the number of cold starts.}
        \label{fig:cold_starts}
    \end{minipage}
\end{figure*}

\subsection{Performance of Pull-Based Scheduling}
Our experimental results show that pull-based scheduling significantly improves several performance metrics over traditional scheduling algorithms. The following analysis details the performance of pull-based scheduling compared to the baseline scheduling algorithms.

\altparagraph{Latency:}
Figure~\ref{fig:results_latency} shows the cumulative distribution function (CDF) of response latencies on OpenLambda using different scheduling algorithms. Pull-based scheduling improves response latencies, as shown by its CDF being the leftmost consistently.
The average response latency in Figure~\ref{fig:results_average_latency} using pull-based scheduling is 481~ms, whereas the contenders score between 565 and 660~ms. Pull-based scheduling reduces the response latencies between 14.9\% and 27.1\%.
We also compare the 90th, 95th and 99th latency percentiles in Figure~\ref{fig:results_tail_latency}. Pull-based scheduling scores lower tail latencies compared to the contenders, particularly at the 99th percentile where pull-based scheduling reduces response latencies by up to 36.4\% on average,
which shows that the performance of pull-based scheduling is stable.
We also analyze the scheduling overhead for each algorithm and find it to be negligible,  ranging from an average of 0.0023 ms for random scheduling to 0.0149 ms for pull-based scheduling.

\altparagraph{Cold starts:}
We evaluate the rate of cold starts in Figure~\ref{fig:cold_starts}, which are a major contributor to high latencies in serverless computing. Pull-based scheduling increases function locality and significantly reduces the number of cold starts compared to the other schedulers. 30\% of requests using pull-based scheduling experience a cold start, whereas requests using other scheduling algorithms experience between 43\% and 59\% cold starts.

\begin{figure*}[!th]
    \centering
    \begin{minipage}{0.49\textwidth}
        \centering
        \includegraphics[width=\textwidth]{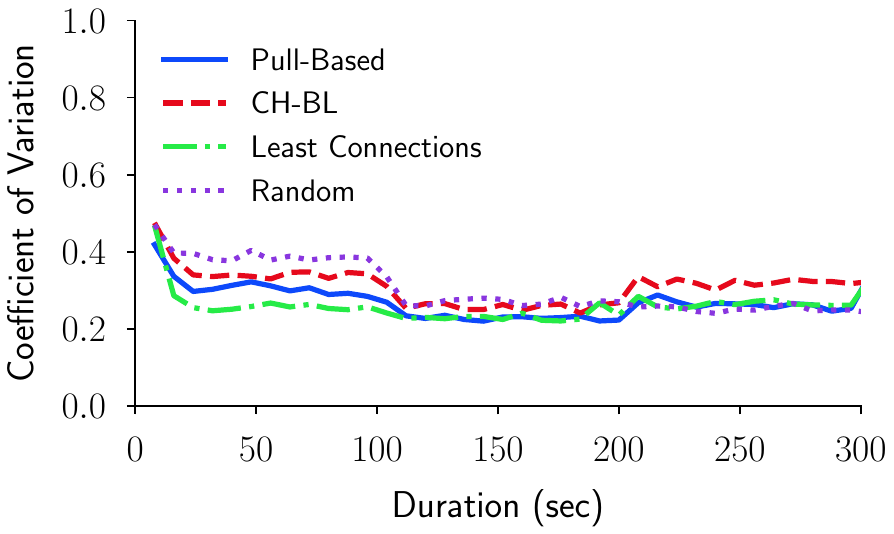}
        \caption{Load imbalance: Coefficient of variation of tasks assigned per second. Pull-based scheduling achieves a comparable level of load balancing as least-connection scheduling.}
        \label{fig:results_load_imbalance}
    \end{minipage}
    \hfill
    \begin{minipage}{0.49\textwidth}
        \centering
        \includegraphics[width=\textwidth]{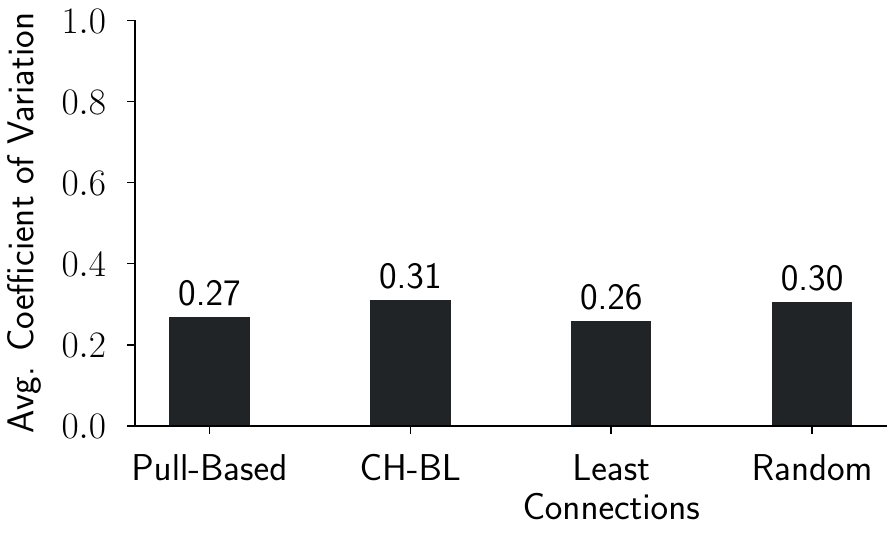}
        \caption{Average load imbalance: Coefficient of variation of tasks assigned per second. Pull-based scheduling balances loads 12.9\% more evenly than consistent hashing with bounded loads.}
        \label{fig:results_average_load_imbalance}
    \end{minipage}
\end{figure*}

\altparagraph{Load imbalance:}
We next evaluate how evenly the scheduling algorithms balance loads. Figure~\ref{fig:results_load_imbalance} shows the coefficient of variation (CV) of tasks assigned per second for the different scheduling algorithms. We compare the average CV in Figure~\ref{fig:results_average_load_imbalance}. Pull-based scheduling (CV: 0.27) achieves a comparable level of load balancing as least-connection scheduling (average CV: 0.26) and balances loads 12.9\% more evenly than consistent hashing with bounded loads (CV: 0.31),
which demonstrates the self-balancing behavior of pull-based scheduling.

\begin{figure*}[!th]
    \centering
    \begin{minipage}{0.49\textwidth}
        \centering
        \includegraphics[width=\columnwidth]{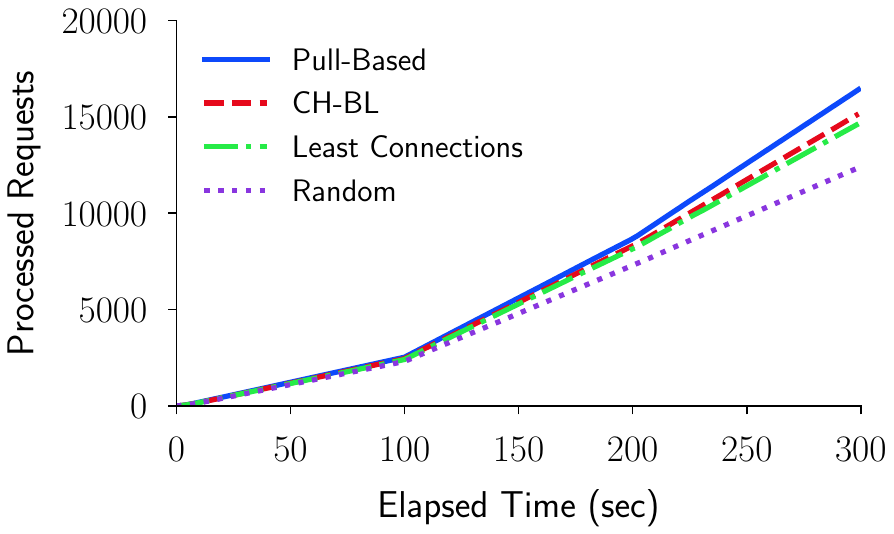}
        \caption{Throughput. Pull-based scheduling increases throughput by up to 32.8\%.}
        \label{fig:results_throughput}
    \end{minipage}
    \hfill
    \begin{minipage}{0.49\textwidth}
        \centering
        \includegraphics[width=\columnwidth]{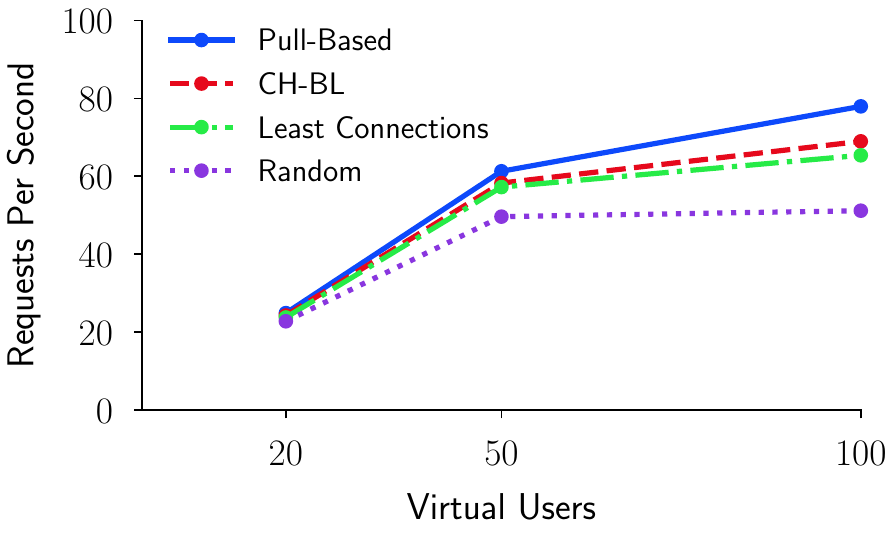}
        \caption{Concurrency. Pull-based scheduling performs best under high concurrency.}
        \label{fig:concurrency}
    \end{minipage}
\end{figure*}

\altparagraph{Throughput:}
Figure~\ref{fig:results_throughput} shows the cumulative number of requests processed over time for each scheduling algorithm. Pull-based scheduling processes an average of 16414 requests, whereas the other scheduling algorithms process between 12361 and 15151 requests, which is an increase in throughput of between 8.3\% and 32.8\%.

\altparagraph{Concurrency:}
We analyze the impact of concurrency on system performance in Figure~\ref{fig:concurrency} by measuring the number of requests processed per second (rps) at 20, 50, and 100 virtual users (VUs). At 20 VUs, all scheduling algorithms show similar levels of performance. At 50 VUs, pull-based scheduling processes 61.3~rps and slightly outperforms consistent hashing with bounded loads, which processes 58.3~rps. At 100 VUs, pull-based scheduling significantly increases throughput with 78~rps, as the other scheduling algorithms process between 51.2 and 69~rps. Overall, pull-based scheduling outperforms the other scheduling algorithms, especially under high concurrency.

\section{Related Work}
\label{sec:related_work}

Scheduling has been an active area of research. For our work, we focus on related works in serverless computing and queuing theory as the most relevant ones. In the following, we position pull-based scheduling within these two areas.

\altparagraph{FaaS scheduling:}
Almost all prior work on FaaS scheduling follows a \textit{push-based} model, where the scheduler tries to select suitable workers for incoming function requests. Examples include the following:
Fuerst and Sharma~\cite{fuerst2022locality} achieve function locality in serverless clusters with a variation of consistent hashing, which randomly updates loads to prevent server overload during bursty workloads.
Abdi et al.~\cite{abdi2023palette} increase data locality in serverless computing by caching results in idle function instances, and introducing a color-based system for pre-defined locality hints that uses consistent hashing for color mappings, which requires additional configuration from developers.
Aumala et al.~\cite{aumala2019beyond} propose a different method to increase data locality using package-aware scheduling, which caches packages at workers and uses consistent hashing to assign requests to workers with preloaded packages.

Our approach fundamentally shifts from this push-based scheduling to \textit{pull-based} scheduling, where idle workers proactively request tasks. Furthermore, unlike most previous work, we do not rely on consistent hashing to achieve locality, thus avoiding its fundamental limitations (see Section~\ref{sec:pull_based_scheduling}).
The closest work to ours is by Kim and Roh~\cite{kim2021scheduling}, who propose predictive pre-warming of containers based on the length of request queues while containers continuously pull tasks for execution, which has similarities to the pull mechanism we propose in this paper. Several other pre-warming techniques have been proposed for serverless computing~\cite{roy2022icebreaker, agarwal2021reinforcement, silva2020prebaking}, but the timing of instance pre-warming and the number of pre-warmed instances required can be inaccurate and incur significant costs, not present in pull-based scheduling.

\altparagraph{Queuing theory:}
In queuing theory, traditional algorithms such as Join-Shortest Queue~\cite{eschenfeldt2018join} or Power-of-d-choices~\cite{hellemans2018power} are push-based.
Lu et. al~\cite{lu2011join} are the first to propose the class of Join-Idle-Queue (JIQ) algorithms, where idle servers inform the scheduler of their readiness for new tasks. JIQ has been successfully applied in large distributed systems to reduce system load and response times compared to push-based scheduling algorithms~\cite{wang2018distributed}. Although pull-based scheduling has received attention in the queuing theory literature, its application to serverless computing has remained largely unexplored.

\section{Conclusion}
\label{sec:conclusion}

In this paper, we proposed a novel pull-based scheduling algorithm for serverless computing. Our approach diverges from traditional scheduling algorithms by allowing idle workers to actively pull tasks to maximize function locality.
Experimental evaluation showed that pull-based scheduling significantly improves response latencies, throughput, and load balancing, while reducing cold starts in serverless computing.

\section{Open Data}
\label{sec:open_data}
We published a replication package~\cite{akbari2025} containing the implementation of pull-based scheduling for OpenLambda and contending scheduling algorithms for comparison. We provided detailed instructions for replicating our experiments, along with scripts and the raw data collected.

\balance
\bibliographystyle{bibliography/IEEEtran}
\bibliography{bibliography/IEEEabrv,bibliography/IEEEbibliography}

\begin{thebibliography}{10}
\providecommand{\url}[1]{#1}
\csname url@samestyle\endcsname
\providecommand{\newblock}{\relax}
\providecommand{\bibinfo}[2]{#2}
\providecommand{\BIBentrySTDinterwordspacing}{\spaceskip=0pt\relax}
\providecommand{\BIBentryALTinterwordstretchfactor}{4}
\providecommand{\BIBentryALTinterwordspacing}{\spaceskip=\fontdimen2\font plus
\BIBentryALTinterwordstretchfactor\fontdimen3\font minus \fontdimen4\font\relax}
\providecommand{\BIBforeignlanguage}[2]{{%
\expandafter\ifx\csname l@#1\endcsname\relax
\typeout{** WARNING: IEEEtran.bst: No hyphenation pattern has been}%
\typeout{** loaded for the language `#1'. Using the pattern for}%
\typeout{** the default language instead.}%
\else
\language=\csname l@#1\endcsname
\fi
#2}}
\providecommand{\BIBdecl}{\relax}
\BIBdecl

\bibitem{aws_lambda}
\BIBentryALTinterwordspacing
Amazon, ``{AWS Lambda},'' 2014. [Online]. Available: \url{https://aws.amazon.com/lambda}
\BIBentrySTDinterwordspacing

\bibitem{google_cloud_functions}
\BIBentryALTinterwordspacing
Google, ``{Cloud Functions},'' 2016. [Online]. Available: \url{https://cloud.google.com/functions}
\BIBentrySTDinterwordspacing

\bibitem{microsoft_azure_functions}
\BIBentryALTinterwordspacing
Microsoft, ``{Azure Functions},'' 2017. [Online]. Available: \url{https://azure.microsoft.com/services/functions}
\BIBentrySTDinterwordspacing

\bibitem{hendrickson2016serverless}
S.~Hendrickson, S.~Sturdevant, T.~Harter, V.~Venkataramani, A.~C. Arpaci-Dusseau, and R.~H. Arpaci-Dusseau, ``Serverless computation with {OpenLambda},'' in \emph{8th USENIX workshop on hot topics in cloud computing (HotCloud 16)}, 2016.

\bibitem{openwhisk}
\BIBentryALTinterwordspacing
``{Apache OpenWhisk},'' 2016. [Online]. Available: \url{https://openwhisk.apache.org}
\BIBentrySTDinterwordspacing

\bibitem{openfaas}
\BIBentryALTinterwordspacing
``{OpenFaaS},'' 2016. [Online]. Available: \url{https://www.openfaas.com}
\BIBentrySTDinterwordspacing

\bibitem{knative}
\BIBentryALTinterwordspacing
``{Knative},'' 2020. [Online]. Available: \url{https://knative.dev}
\BIBentrySTDinterwordspacing

\bibitem{ali2020batch}
A.~Ali, R.~Pinciroli, F.~Yan, and E.~Smirni, ``Batch: Machine learning inference serving on serverless platforms with adaptive batching,'' in \emph{SC20: International Conference for High Performance Computing, Networking, Storage and Analysis}.\hskip 1em plus 0.5em minus 0.4em\relax IEEE, 2020, pp. 1--15.

\bibitem{fouladi2019laptop}
S.~Fouladi, F.~Romero, D.~Iter, Q.~Li, S.~Chatterjee, C.~Kozyrakis, M.~Zaharia, and K.~Winstein, ``From laptop to lambda: Outsourcing everyday jobs to thousands of transient functional containers,'' in \emph{2019 USENIX annual technical conference (USENIX ATC 19)}, 2019, pp. 475--488.

\bibitem{li2022funcx}
Z.~Li, R.~Chard, Y.~Babuji, B.~Galewsky, T.~J. Skluzacek, K.~Nagaitsev, A.~Woodard, B.~Blaiszik, J.~Bryan, D.~S. Katz \emph{et~al.}, ``Funcx: Federated function as a service for science,'' \emph{IEEE Transactions on Parallel and Distributed Systems}, vol.~33, no.~12, pp. 4948--4963, 2022.

\bibitem{mahgoub2022wisefuse}
A.~Mahgoub, E.~B. Yi, K.~Shankar, E.~Minocha, S.~Elnikety, S.~Bagchi, and S.~Chaterji, ``Wisefuse: Workload characterization and dag transformation for serverless workflows,'' \emph{Proceedings of the ACM on Measurement and Analysis of Computing Systems}, vol.~6, no.~2, pp. 1--28, 2022.

\bibitem{zhangFasterCheaperServerless2021}
Y.~Zhang, {\'I}.~Goiri, G.~I. Chaudhry, R.~Fonseca, S.~Elnikety, C.~Delimitrou, and R.~Bianchini, ``Faster and {{Cheaper Serverless Computing}} on {{Harvested Resources}},'' in \emph{Proceedings of the {{ACM SIGOPS}} 28th {{Symposium}} on {{Operating Systems Principles}}}.\hskip 1em plus 0.5em minus 0.4em\relax Virtual Event Germany: ACM, 2021, pp. 724--739.

\bibitem{silva2020prebaking}
P.~Silva, D.~Fireman, and T.~E. Pereira, ``Prebaking functions to warm the serverless cold start,'' in \emph{Proceedings of the 21st International Middleware Conference}, 2020, pp. 1--13.

\bibitem{fuerst2022locality}
A.~Fuerst and P.~Sharma, ``Locality-aware load-balancing for serverless clusters,'' in \emph{Proceedings of the 31st International Symposium on High-Performance Parallel and Distributed Computing}, 2022, pp. 227--239.

\bibitem{shahrad2020serverless}
M.~Shahrad, R.~Fonseca, I.~Goiri, G.~Chaudhry, P.~Batum, J.~Cooke, E.~Laureano, C.~Tresness, M.~Russinovich, and R.~Bianchini, ``Serverless in the wild: Characterizing and optimizing the serverless workload at a large cloud provider,'' in \emph{2020 USENIX annual technical conference (USENIX ATC 20)}, 2020, pp. 205--218.

\bibitem{wang2018peeking}
L.~Wang, M.~Li, Y.~Zhang, T.~Ristenpart, and M.~Swift, ``Peeking behind the curtains of serverless platforms,'' in \emph{2018 USENIX annual technical conference (USENIX ATC 18)}, 2018, pp. 133--146.

\bibitem{hellemans2018power}
T.~Hellemans and B.~Van~Houdt, ``On the power-of-d-choices with least loaded server selection,'' \emph{Proceedings of the ACM on Measurement and Analysis of Computing Systems}, vol.~2, no.~2, pp. 1--22, 2018.

\bibitem{abdi2023palette}
M.~Abdi, S.~Ginzburg, X.~C. Lin, J.~Faleiro, G.~I. Chaudhry, I.~Goiri, R.~Bianchini, D.~S. Berger, and R.~Fonseca, ``Palette load balancing: Locality hints for serverless functions,'' in \emph{Proceedings of the Eighteenth European Conference on Computer Systems}, 2023, pp. 365--380.

\bibitem{aumala2019beyond}
\BIBentryALTinterwordspacing
G.~Aumala, E.~Boza, L.~Ortiz-Avil{\'e}s, G.~Totoy, and C.~Abad, ``Beyond load balancing: Package-aware scheduling for serverless platforms,'' in \emph{2019 19th IEEE/ACM International Symposium on Cluster, Cloud and Grid Computing (CCGRID)}.\hskip 1em plus 0.5em minus 0.4em\relax IEEE, 2019, pp. 282--291. [Online]. Available: \url{https://github.com/disel-espol/olscheduler}
\BIBentrySTDinterwordspacing

\bibitem{lu2011join}
Y.~Lu, Q.~Xie, G.~Kliot, A.~Geller, J.~R. Larus, and A.~Greenberg, ``Join-idle-queue: A novel load balancing algorithm for dynamically scalable web services,'' \emph{Performance Evaluation}, vol.~68, no.~11, pp. 1056--1071, 2011.

\bibitem{wang2018distributed}
C.~Wang, C.~Feng, and J.~Cheng, ``Distributed join-the-idle-queue for low latency cloud services,'' \emph{IEEE/ACM Transactions on Networking}, vol.~26, no.~5, pp. 2309--2319, 2018.

\bibitem{kimFunctionbenchSuiteWorkloads2019}
J.~Kim and K.~Lee, ``{{FunctionBench}}: {{A}} suite of workloads for serverless cloud function service,'' in \emph{2019 {{IEEE}} 12th International Conference on Cloud Computing ({{CLOUD}})}.\hskip 1em plus 0.5em minus 0.4em\relax IEEE, 2019, pp. 502--504.

\bibitem{agarwal2021reinforcement}
S.~Agarwal, M.~A. Rodriguez, and R.~Buyya, ``A reinforcement learning approach to reduce serverless function cold start frequency,'' in \emph{2021 IEEE/ACM 21st International Symposium on Cluster, Cloud and Internet Computing (CCGrid)}.\hskip 1em plus 0.5em minus 0.4em\relax IEEE, 2021, pp. 797--803.

\bibitem{kim2021scheduling}
D.~K. Kim and H.-G. Roh, ``Scheduling containers rather than functions for function-as-a-service,'' in \emph{2021 IEEE/ACM 21st International Symposium on Cluster, Cloud and Internet Computing (CCGrid)}.\hskip 1em plus 0.5em minus 0.4em\relax IEEE, 2021, pp. 465--474.

\bibitem{roy2022icebreaker}
R.~B. Roy, T.~Patel, and D.~Tiwari, ``Icebreaker: Warming serverless functions better with heterogeneity,'' in \emph{Proceedings of the 27th ACM International Conference on Architectural Support for Programming Languages and Operating Systems}, 2022, pp. 753--767.

\bibitem{mirrokni2018consistent}
V.~Mirrokni, M.~Thorup, and M.~Zadimoghaddam, ``Consistent hashing with bounded loads,'' in \emph{Proceedings of the Twenty-Ninth Annual ACM-SIAM Symposium on Discrete Algorithms}.\hskip 1em plus 0.5em minus 0.4em\relax SIAM, 2018, pp. 587--604.

\bibitem{chen2021revisiting}
J.~Chen, B.~Coleman, and A.~Shrivastava, ``Revisiting consistent hashing with bounded loads,'' in \emph{Proceedings of the AAAI Conference on Artificial Intelligence}, vol.~35, no.~5, 2021, pp. 3976--3983.

\bibitem{schirmer2023night}
T.~Schirmer, N.~Japke, S.~Greten, T.~Pfandzelter, and D.~Bermbach, ``The night shift: Understanding performance variability of cloud serverless platforms,'' in \emph{Proceedings of the 1st Workshop on SErverless Systems, Applications and MEthodologies}, 2023, pp. 27--33.

\bibitem{k6}
\BIBentryALTinterwordspacing
Grafana, ``{k6},'' 2016. [Online]. Available: \url{https://k6.io}
\BIBentrySTDinterwordspacing

\bibitem{eschenfeldt2018join}
P.~Eschenfeldt and D.~Gamarnik, ``Join the shortest queue with many servers. the heavy-traffic asymptotics,'' \emph{Mathematics of Operations Research}, vol.~43, no.~3, pp. 867--886, 2018.

\bibitem{akbari2025}
\BIBentryALTinterwordspacing
S.~Akbari and M.~Hauswirth, ``Replication package for {Hiku}: Pull-based scheduling for serverless computing, v1.0,'' 2025, {DOI}: 10.5281/zenodo.14906185.
\BIBentrySTDinterwordspacing

\end{thebibliography}

\end{document}